\documentclass{article}
\usepackage[a4paper, margin=0.7in]{geometry}
\usepackage{multicol}
\usepackage{amsmath}
\usepackage{graphicx}
\usepackage{caption}
\usepackage{subcaption}
\usepackage{float}
\usepackage{enumitem}
\usepackage[font=small,labelfont=bf]{caption}
\title{\textbf{Determining size of an optically trapped particle via modulated Raman spectroscopy}}
\author{Rohin Sharma}

\author{Rohin Sharma\textsuperscript{1,†}, Anusa Thapa\textsuperscript{1}, Rijan Maharjan\textsuperscript{1}, Ashim Dhakal\textsuperscript{1,*}}

\begin{document}
\maketitle

\noindent
\emph{\textsuperscript{1} Biophotonics Lab, Phutung Research Institute, PO Box 12335, Kathmandu 44600, Nepal}\\
\noindent
\\
† Corresponding email: rohin.sharma@pinstitute.org\\
* Email: ashim.dhakal@pinstitute.org
\paragraph{}
\paragraph{}
\textbf{Abstract:}
The average Raman signal power obtained in a modulated optical trap is dependent on the Brownian motion - therefore hydrodynamic properties of the trapped particle. Hence, in addition to the molecular properties obtained from the Raman signal, it is possible to study hydrodynamic properties (e.g. size) of the particle by analyzing the change in the average Raman power as a function of modulation frequency. Our results, based on the over-damped Langevin equation, show that several minimas exist for the Raman signal at unique modulating frequencies for a given particle size and signal acquisition time. In typical experimental conditions, such minimas can be as low as $\sim$50\% of the Raman signal in an unmodulated trap.
\vspace{5mm}
\hrule
\begin{multicols}{2}

\section{Introduction}
\paragraph{}
Since the seminal report of optical trapping and manipulation of micron-sized particles by exploiting the intensity gradient of a highly focused laser beam \cite{PhysRevLett.24.156}, there have been numerous studies of utilizing this technique to manipulate and study the properties of single molecules, bio molecules and single cells~\cite{smith1996overstretching, hyperstretchingDNA, block1990bead, jagannathan2013protein, ashkin1987optical, Ashkin1987}. The combination of optical trapping and Raman spectroscopy has been a powerful tool for the molecular characterization and identification of micro-particles with high specificity~\cite{thurn1984raman}. This combination has been used in studying aerosol particles~\cite{SCHWEIGER1990483}, gas bubbles~\cite{lankers1997raman}, living organelles~\cite{B108744B}, and cells~\cite{Ashkin1987, xie2002near}, allowing us to gain insights into the molecular properties that would otherwise be inaccessible (such as in sub-cellular level).

The optical trap can be approximated as a harmonic potential with a trap stiffness, $\kappa$. Therefore there have been numerous attempts of calibrating $\kappa$ through various methods such as by measuring the displacement produced by a known external force~\cite{kuo1993force, wuite2000integrated}, or by analysing the force of thermal fluctuations by the power spectrum analysis~\cite{berg2004power}. A very practical method of calibrating $\kappa$ is by periodically modulating the trap stiffness achieved by modulating the optical power of the trapping source~\cite{deng2007brownian}. The effects of such modulated optical trap on Brownian motion and the position variance of the trapped particle were studied both theoretically and experimentally along with the report of errors it can produce in force measurements in Ref.~\cite{deng2007brownian}. This method has subsequently been applied to perform quantitative measurement with bio-molecules ~\cite{Horst:08}.

Because the trap modulation directly modulates the harmonic potential, understanding the effects of trap modulation on the motion of Brownian particle is very important in optical traps, and has gained considerable interest in recent decades. While there have been studies on the dynamics of Brownian particle in the modulated harmonic trap~\cite{deng2007brownian, PhysRevLett.93.160602, PhysRevLett.99.010601, PhysRevE.77.051107, PhysRevE.81.051123, zerbe1994brownian}, the effects of such modulation on the Raman signal of the trapped particle has not yet been explored. Since Raman spectroscopy does not distinguish materials of similar chemical composition but with different macro-level physical properties (eg. size, shape), identifying the response of trap modulation on the Raman signal would, in addition, allow investigations of hydrodynamic and optical properties of the trapped particle. Based on the theoretical formulation of the position variance of optically trapped particle shown in Ref.~\cite{deng2007brownian}, our work here will theoretically estimate the variations in Raman power of the trapped particle as a function of modulation frequency.

In section 2, we discuss the theoretical background and introduce the expression for average Raman power. In section 3, we discuss the findings of our result.

\section{Theoretical formulation}
\paragraph{}
Since an optically trapped micron sized particle in a viscous medium exhibits Brownian motion, the particle has a natural tendency to diffuse away from the equilibrium position. The average time the particle will stay in the equilibrium position is determined by the the inverse of its \textit{roll-off frequency, $\omega_o$}; defined as the ratio of \text{trap-stiffness}, $\kappa$ and \text{drag coefficient}, $\gamma$. As $\kappa$ is directly proportional to the laser power, a sinusoidally modulated laser power results in the modulation of $\kappa$. So, the equation of motion given by the Langevin equation is

\begin{equation} \label{eom}
    \dot{x}(t) + \frac{\kappa}{\gamma}\left[1 + \epsilon cos(\omega t)\right]x(t) = \sqrt{2D} \eta(t).
\end{equation}

Here, $x(t)$ is the trajectory of the Brownian particle, $\omega$ is the modulation frequency, $\epsilon$ is the modulation depth, $-\kappa x(t)$ is the harmonic force from the trap, and $D = k_BT/\gamma$, is the Einstein's equation relating diffusion constant with the Boltzmann energy and drag coefficient. For a spherical particle of radius $r$ trapped in a medium of density $\rho$ and kinematic viscosity $\nu$, the drag coefficient $\gamma$ is approximated by the Stokes' law as $\gamma = 6\pi \rho \nu r$. $\sqrt{2D}\eta(t)$, is a stochastic Gaussian process that represents Brownian forces at absolute temperature $T$ such that for all $t$ and $t'$:
\begin{equation}
\begin{split}
    \langle \eta (t) \rangle = 0; \\ 
    \langle \eta(t) \eta(t') \rangle = \delta (t - t').
\end{split}
\end{equation}
Following the derivation of It\^{o}'s lemma \cite{gardiner1985handbook, itostochastic}, the position variance of the trapped particle satisfies the equation (see Appendix A for the derivation)

\begin{equation} \label{de}
    \dot{\sigma}^2_{xx}(t) = 2\left[D - \omega_o (1 + \epsilon cos \omega t)\sigma_{xx}^2(t)\right].
\end{equation}

From the equipartition theorem the initial position variance in the $x$ direction takes the equilibrium value of

\begin{equation} \label{sigma_o}
    \sigma_{xx}^2(0) = \frac{k_BT}{\kappa}.
\end{equation}

\begin{figure}[H]
    \centering
    \includegraphics[width = 8.0 cm]{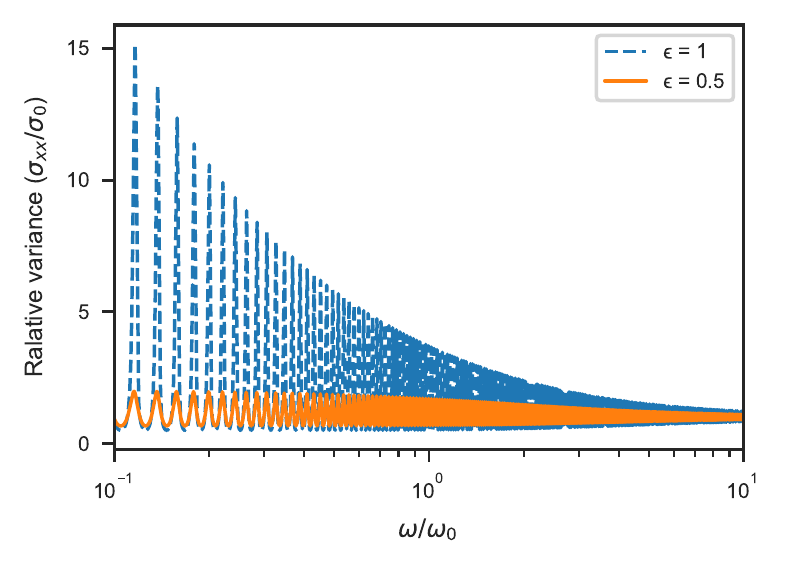}
    \caption{Position variance as a function of modulation frequency, $\omega$ at two different modulation depths, $\epsilon = 1$ and $0.5$ for an integration time of $300$ oscillation cycles. The transition from high to low-frequency occurs at the roll-off frequency, $\omega_o$.}
    \label{variance_plot}
\end{figure}

The solution of Eq.~\ref{de} with the initial position variance $\sigma_{xx}^2(0)$ is the position variance of the trapped particle. Writing the ratio of $\sigma_{xx}^2(t) / \sigma_{xx}^2(0)$ as $V(t)$ we have
\begin{multline} \label{var_eq}
    V(t) = 1 - 2\omega_o \epsilon \exp\left[-2\omega_o \left(t + \frac{\epsilon}{\omega}sin \omega t\right)\right] \\
    \times \int_0^t \exp\left[{2\omega_o\left(\tau + \frac{\epsilon}{\omega} sin \omega \tau\right)}\right]cos \omega \tau d\tau.
\end{multline}
This equation gives insights as to how the modulation frequency affects the position variance of the trapped particle. From the simulation of Eq.~(\ref{var_eq}) in Fig.~\ref{variance_plot}, we see that the position variance approaches the equilibrium value, $\sigma_{xx}^2(0)$ in the high modulation frequency region, where as in the low frequency region, the particle will have enough time to diffuse away from the trapping potential, showing significant position variance.

\subsection*{Raman Power of the trapped particle}
\paragraph{}
We are interested to know the nature of Raman power that can be collected from the optically trapped particle as a function of modulation frequency.
For a trapped particle at position, $x(t)$ the Raman signal emitted from the particle with the Raman cross-section, $\sigma_R$ is
\begin{equation}
    P(x) = \sigma_R I_o \exp\left[-2\frac{x^2(t)}{\sigma^2_{xx}(t)}\right],
\end{equation}
where $I_o$ is the intensity of the laser. Assuming the same excitation and collection coefficient, the normalized Raman power is
\begin{equation} \label{Norm_Raman}
    R(x) = \frac{P(x)}{\sigma_R I_o} = \exp\left[-4\frac{x^2(t)}{w^2(z)}\right],
\end{equation}
Writing $w(z)$ in terms of the position variance at the equilibrium $\sigma_{xx}^2(0)$
\begin{equation} \label{w}
    w^2(z) = \xi \sigma_{xx}^2(0),
\end{equation}
 where $\xi$ is the conversion factor, the average Raman power of modulated Raman signal will be (see Appendix B for an alternate derivation using It\^{o}'s lemma \cite{gardiner1985handbook, itostochastic} for stochastic processes)
\begin{equation} \label{R_mod}
    R_{mod} = \langle R(x) \rangle = \sqrt{\frac{\xi}{8V(t) + \xi}}.
\end{equation}
 For unmodulated case, $V(t) = 1$
\begin{equation} \label{R_unmod}
    R_{unmod} = \sqrt{\frac{\xi}{8 + \xi}},
\end{equation}
which is the change in Raman signal due to diffusion. Therefore,
\begin{equation} \label{Raman}
    \frac{R_{mod}}{R_{unmod}} = \sqrt{\frac{8 + \xi}{8 V(t) + \xi}}.
\end{equation}
Using Eq.~(\ref{var_eq}) and choosing the value of the conversion factor, $\xi$ according to the source Gaussian beam, we can analyse Eq.~(\ref{Raman}) to gain some insights into the nature of Raman signal with respect to the modulating frequency.

\section{Results and discussion}
\begin{figure}[H]
    \centering
    \includegraphics[width = 8cm]{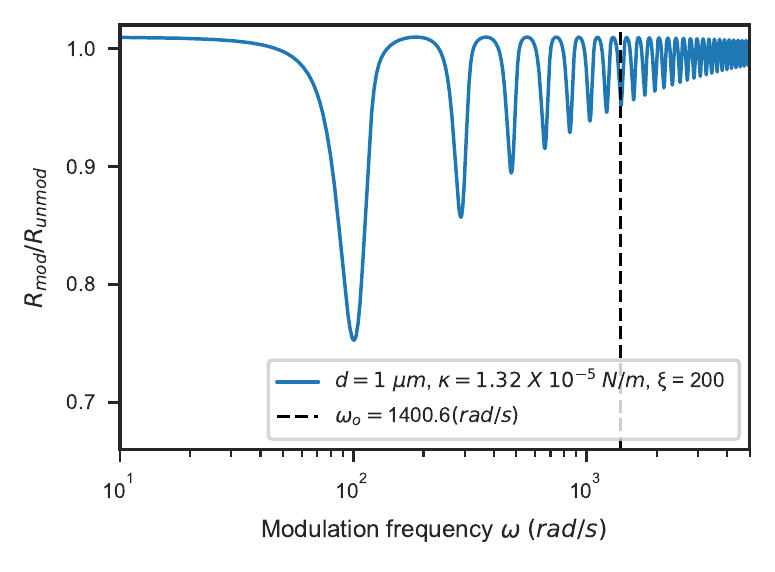}
    \caption{Raman power variance of a one micron trapped particle plotted as a function of modulation frequency, $\omega$ from $10$ $rad/s$ to $5000$ $rad/s$ for trap-stiffness, $\kappa = 1.32 \times 10^{-5}$ $N/m$ and conversion factor, $\xi = 200$. The vertical line denote the roll-off frequency at $\omega_o = 1400.6$ $rad/s$ which defines the high-frequency and low-frequency region for the trap modulation.}
    \label{fig2}
\end{figure}

\begin{figure}[H]
    \centering
    \includegraphics[width = 8cm]{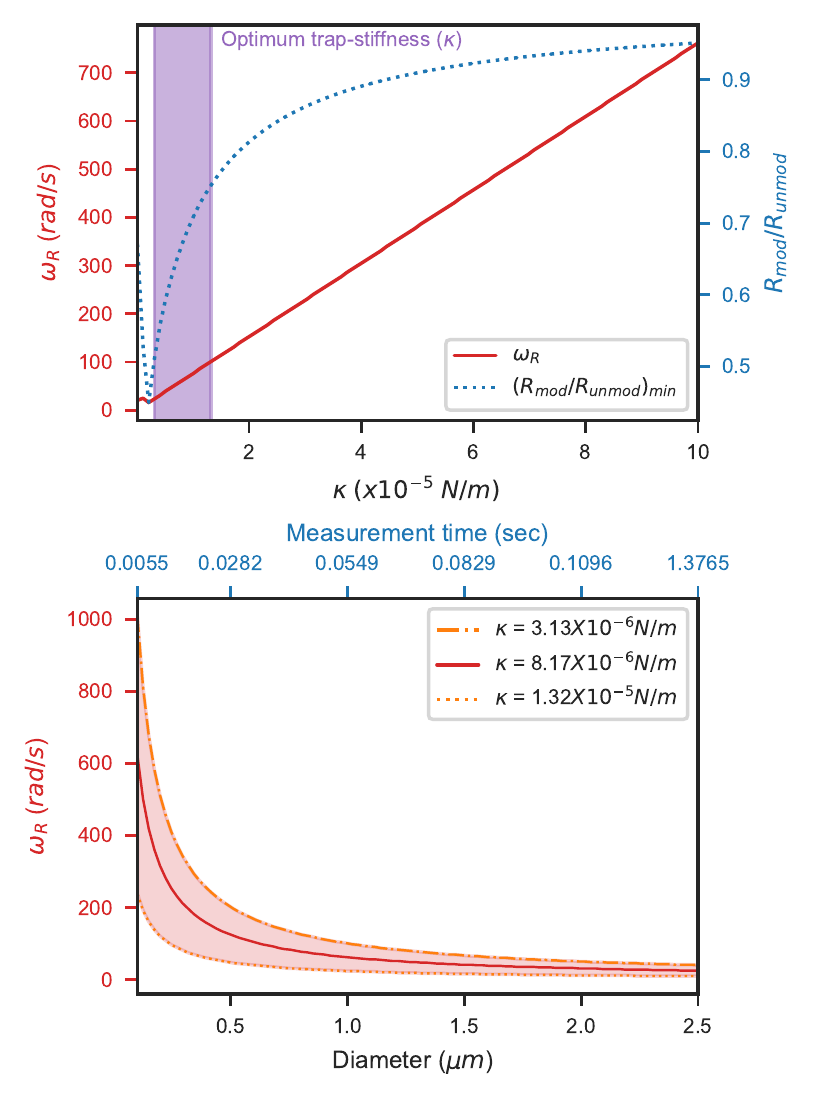}
    \caption[8pt]{(a) The dependence of maximum Raman power variance, $\left(\frac{R_{mod}}{R_{unmod}}\right)_{min}$ and the frequency corresponding to the maximum Raman power variance, $\omega_R$ on the trap stiffness, $\kappa$ for a one micron diameter particle in the modulation frequency range of $10$ $rad/s$ to $1500$ $rad/s$. The average optimum trap-stiffness is $8.17 \times 10^{-6}$ $N/m$, where about 35\% Raman power variation is seen. The discontinuities seen below $\kappa = 1.32 \times$ $10^{-5}$ $N/m$ is due to the theoretical limits of the approximation presented in Appendix B. (b) The profile of modulation frequencies showing maximum Raman power variance with respect to the diameter of the particle along with the corresponding measurement time (shows at top axis) for the optimized trap-stiffness.} 
 \label{fig3}
\end{figure}

\paragraph{}
For an optical trap with a source wavelength, $\lambda = 785nm$, a high Numerical Aperture objective lens of $NA = 1$, the Gaussian beam waist radius at the focus $w(z = 0)$ is $250$ $nm$. From the simulation of Eq.~(\ref{Raman}) for a $1$ $\mu m$ particle trapped in an aqueous  medium with $\rho = 10^3$ $kg/m^3$ as the density of water at room temperature and $\nu = 10^{-6}$ $m^2/s$ as the kinematic friction coefficient of water shown in Fig.~\ref{fig2}, we see that the \textit{maximum Raman power variance}, ${(R_{mod}/R_{unmod})}_{min}$, is seen at a particular modulation frequency, $\omega_R$ in the low frequency region ($\omega < \omega_o$) due to the increased position variance of the particle in the low-frequency region, while in the high frequency region the Raman power approaches the value for an unmodulated case given by Eq.~(\ref{R_unmod}). Since from Eq.~(\ref{Raman}) we know that the Raman power variance depends on the trap-stiffness $\kappa$ and drag coefficient $\gamma$ (determined by the the friction coefficient of the medium and particle diameter), and also since the conversion factor $\xi$ is dependent on the trap stiffness, it is necessary to first calibrate the trap stiffness of the optical trap. We first analyse Eq.~(\ref{Raman}) over a range of trap stiffness to optimize $\kappa$ that show maximum average Raman power variance for a $1$ $\mu m$ trapped particle with the measurement time $t$ as that for 300 cycles of oscillation shown in Fig.~\ref{fig3}a. Then for the optimised $\kappa$, we show the dependence of $\omega_R$ with the diameter, $d$ of the trapped particle. 

From Fig.~\ref{fig3}b we see that for a given particle size, the maximum Raman power variance occurs at a unique modulation frequency, $\omega_R$ for the measurement time of a 300 oscillation cycles. Hence it is possible to calibrate the size of the trapped particle according to the modulating frequency at which the maximum Raman power variance is seen. The dependence of Raman power variance on the diameter of the particle for a given $\kappa$ was insignificant (data not shown).

\begin{figure}[H]
    \centering
    \includegraphics[width = 8cm]{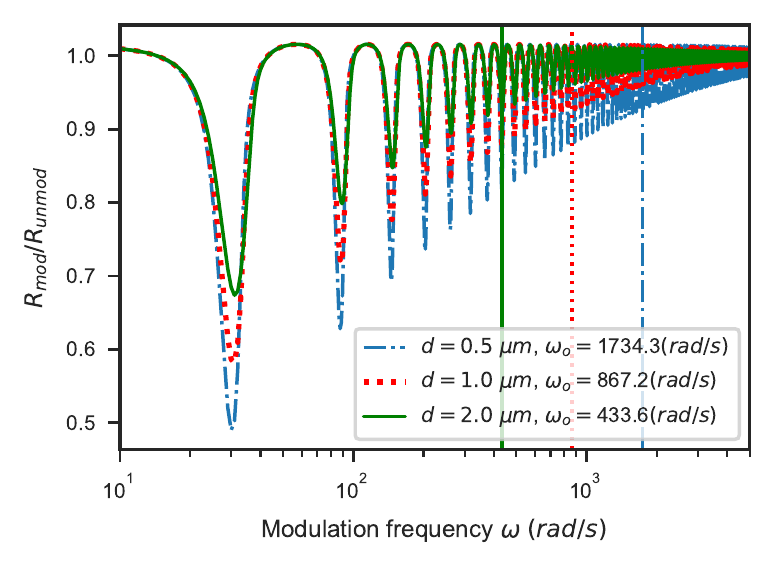}
    \caption{Raman power fluctuations for three different diameters $0.5$ $\mu m$, $1.0$ $\mu m$ and $2.0$ $\mu m$ for $\kappa = 8.17 \times 10^{-6}$ $N/m$, showing fluctuations of approximately $50\%$, $41\%$ and $37\%$ respectively.}
    \label{fig4}
\end{figure}

The particle size calibration can also be done by solely studying the Raman power. If the measurement is done at a constant measurement time for particles of different sizes in a given trap-stiffness, then the Raman power variances for different particles will be seen at the same modulation frequency, but the values of such fluctuations will be different according to the size of the particle as shown in Fig.~\ref{fig4}. In such a case, the size of the trapped particle can be calibrated by studying the Raman signal fluctuations in the modulated optical trap.

To conclude, in addition to molecular identification and characterization offered by Raman spectroscopy, the size of the trapped particle can be investigated in a single experiment by measuring the average Raman power as a function of the modulation frequency relative to the roll-off frequency in the low frequency regime.

\section{Acknowledgement}
\paragraph{}
This work was partially supported by TWAS/SIDA/UNESCO under Grants 18-013RG/Phys/AS\_I and 21-334 RG/PHYS/AS\_G.

\end{multicols}
\section{Appendix}
\subsection*{Appendix A}

The overdamped equation of motion of the trapped particle is 
\begin{equation} \label{Le}
    \dot{x}(t) + \frac{\kappa}{\gamma}\left[1 + \epsilon cos(\omega t)\right]x(t) = \sqrt{2D} \eta(t).
\end{equation}
where $\kappa/\gamma = \omega_o$ is the particle's roll-off frequency. Comparing Eq.~(\ref{Le}) with the stochastic differential equation of the form
\begin{equation}
    dX_t = \mu_t dt + \sigma_t dB_t,
\end{equation}
we have $\mu = -\omega_o(1 + \epsilon cos\omega t)x_t$ and $\sigma_t = \sqrt{2D}$. Here $B_t$ is a Wiener process \cite{gardiner1985handbook}. To calculate the position variance of the Brownian motion described by Eq.~(\ref{Le}) we apply It{\^o}'s lemma~\cite{itostochastic} to calculate the mean of $x^2$. Taking $f = x^2$, $df$ is calculated as
\begin{equation} \label{sde}
    df = \left(\frac{\partial f}{\partial t} + \mu_t \frac{\partial f}{\partial x} + \frac{\sigma^2_t}{2}\frac{\partial^2f}{\partial x^2}\right)dt + \sigma_t \frac{\partial f }{\partial x}dB_t.
\end{equation}
Here $\langle \frac{\partial f}{\partial t}\rangle = 0$, $\langle \mu_t \frac{\partial f}{\partial x}\rangle = -2 \omega_o (1 + \epsilon cos\omega t) \sigma^2_{xx}(t)$ and $\langle \frac{\sigma_t^2}{2} \frac{\partial ^2 f}{\partial x^2} \rangle = 2D$ and $\langle \sigma_t \frac{\partial f}{\partial x} dB_t \rangle = 0$ as $\langle xdB_t \rangle = 0$.
So the position variance is
\begin{equation}
\begin{split}
    d\langle x^2 \rangle = d\sigma^2_{xx}(t) = -2 \omega_o (1 + \epsilon cos \omega t) \sigma^2_{xx}(t)dt + 2Ddt.\\
    \dot{\sigma}^2_{xx}(t) = 2\left[D - \omega_o(1 + \epsilon cos \omega t)\sigma^2_{xx}(t)\right].
\end{split}
\end{equation}

\subsection*{Appendix B}
Here we present an alternate derivation of the Raman signal for a modulated Gaussian beam using the stochastic differential equation.\\
The Taylor's expansion of the normalized Raman power Eq.~(\ref{Norm_Raman}) is:
\begin{equation}
    R(x) = 1 - \frac{4x^2}{w^2(z)} + \frac{192 w^4(z) x^4}{4! \times w^{10}} - ...
\end{equation}

Taking the first two terms, we write $f = 1 - \frac{4x^2}{w^2(z)}$. Since $\mu_t = -\omega_o(1 + \epsilon cos \omega t)x_t$ and $\sigma_t = \sqrt{2D}$, $\langle \mu_t \frac{\partial f}{\partial x} \rangle = 8\omega_o(1 + \epsilon cos \omega t) \frac{\sigma^2_{xx}(t)}{w^2(t)}$, $\langle \frac{\sigma^2_t \partial^2 f}{2\partial x^2} \rangle = -8\frac{D}{w^2(z)}$ and $\langle \sigma_t \frac{\partial f}{\partial x} dB_t \rangle = 0$ as $\langle xdB_t \rangle = 0$. Using Eq.~(\ref{sde}), we have

\begin{equation}
    d\left\langle 1 - 4\frac{x^2}{w^2(z)} \right\rangle = \frac{8}{w^2(z)}\left[\omega_o \sigma^2_{xx}(t)(1 + \epsilon cos \omega t) - D\right]dt.
\end{equation}
Using equations Eq.~(\ref{var_eq}) and Eq.~(\ref{w}) to express in terms of $V(t)$ and $\xi$ we have,
\begin{equation} \label{approx}
    \frac{d}{dt}\left\langle R'(x) \right\rangle = \frac{8 \omega_o}{\xi}\left[(1 + \epsilon cos \omega t) V(t) - 1\right].
\end{equation}
Now to compare this expression with our equation of the normalized Raman power for the modulated case, we take the time derivative of Eq.~(\ref{R_mod}).
\begin{equation} \label{exact}
    \frac{d}{dt} \langle R(x) \rangle = \frac{8 \omega_o}{\xi}\left[(1 + \epsilon cos \omega t) V(t) - 1\right] \left(\frac{\xi}{\xi + 8V(t)}\right)^{\frac{3}{2}}.
\end{equation}
Since Eq.~(\ref{approx}) and Eq.~(\ref{exact}) differs by a factor $\left(\frac{\xi}{\xi + 8V(t)}\right)^{\frac{3}{2}}$, for an optical trap system with the $\xi$ value such that, the factor $\left(\frac{\xi}{\xi + 8V(t)}\right)^{\frac{3}{2}} \sim 1$, Eq.~(\ref{Raman}) would be a good approximation to estimate the fluctuations of the average Raman signal for the modulated case.

\bibliography{ref}
\bibliographystyle{ieeetr}

\end{document}